\documentclass[proceedings]{stacs}
\stacsheading{2009}{685--696}{Freiburg}
\firstpageno{685}

\begin{document}
\def\maj{\mathbf{MAJ}}
\def\brec{\mathbf{BRecMAJ}}
\def\urec{\mathbf{URecMAJ}}

\title[LP Bound via Clique Constraints]
{A Stronger LP Bound for Formula Size Lower Bounds via Clique Constraints}
\author[UT]{K. Ueno}{Kenya Ueno}
\address[UT]{Department of Computer Science, 
Graduate School of Information Science and Technology,
The University of Tokyo}
\email{kenya@is.s.u-tokyo.ac.jp}

\keywords{Computational and Structural Complexity}
\subjclass{F.1.1 Models of Computation}

\begin{abstract}
\noindent We introduce a new technique proving formula size lower bounds based on the linear programming bound originally introduced by Karchmer, Kushilevitz and Nisan~\cite{KKN95} and the theory of stable set polytope. We apply it to majority functions and prove their formula size lower bounds improved from the classical result of Khrapchenko~\cite{Khrapchenko71}. Moreover, we introduce a notion of unbalanced recursive ternary majority functions motivated by a decomposition theory of monotone self-dual functions and give integrally matching upper and lower bounds of their formula size. We also show monotone formula size lower bounds of balanced recursive ternary majority functions improved from the quantum adversary bound of Laplante, Lee and Szegedy~\cite{LLS06}.
\end{abstract}

\maketitle
\section{Introduction}
Proving formula size lower bounds is a fundamental problem in complexity theory 
and also an extremely tough problem to resolve.
A super-polynomial lower bound of a function in $\mathbf{NP}$ implies 
$\mathbf{NC}_1 \neq \mathbf{NP}$.
There are a lot of techniques to prove formula size lower bounds,
e.g. \cite{Hastad98,HLS07,KKN95,Khrapchenko71,Koutsoupias93,LLS06,Lee07}.
Laptente, Lee and Szegedy~\cite{LLS06} introduced a technique 
based on the quantum adversary method~\cite{Ambainis02} and 
gave a comparison with known techniques.
In particular, they showed that their technique subsumes several known techniques 
such as Khrapchenko~\cite{Khrapchenko71} and its extension~\cite{Koutsoupias93}.
The current best formula size lower bound is $n^{3-o(1)}$ by H{\aa}stad~\cite{Hastad98} and
a key lemma used in the proof is also subsumed 
by the quantum adversary bound~\cite{LLS06}.
Karchmer, Kushilevitz and Nisan~\cite{KKN95} introduced a technique 
proving formula size lower bounds called the linear programming (or LP) bound 
and showed that it cannot prove a lower bound
larger than $4 n^2$ for non-monotone formula size in general.
Lee~\cite{Lee07} proved that the LP bound~\cite{KKN95}
subsumes the quantum adversary bound~\cite{LLS06} and
H{\o}yer, Lee and \v{S}palek~\cite{HLS07} introduced a stronger version of 
the quantum adversary bound.

Motivated by the result of Lee~\cite{Lee07}, we devise a stronger version
of the LP bound by using an idea from the theory of stable set polytope, 
known as clique constraints~\cite{Padberg73}.
Suggesting a stronger technique compared to the original LP bound~\cite{KKN95}
has possibilities to improve the best formula size lower bound 
because it subsumes many techniques including
the key lemma of H{\aa}stad~\cite{Hastad98}.
Moreover, our technique has various possibilities of extensions such as
rank constraints discussed in Section~\ref{brec_section} and 
orthonormal constraints~\cite{GLS88}, each of which subsume clique constraints.
Due to this extendability, it is difficult to show the limitation of our new technique.

To study the relative strength of our technique, we apply it to some families of Boolean functions.
For each family, we have distinct motivation to investigate their formula size.
Three kinds of Boolean functions treated in this paper are defined as follows.
All of them are called monotone self-dual Boolean functions defined in the next section.
\begin{definition}
A majority function $\maj_{2l+1} : \{0,1\}^{2l+1} \mapsto \{0,1\}$ outputs $1$ 
if the number of 1's in the input bits is greater than or equal to $l+1$ and $0$ otherwise.
We define unbalanced recursive ternary majority functions
$\urec_3^h : \{0,1\}^{2h+1} \mapsto \{0,1\}$ as 
\[
\urec_3^h (x_1, \cdots, x_{2h+1}) = \maj_3 (
\urec_3^{h-1}(x_{1}, \cdots, x_{2h-1}), x_{2h}, x_{2h+1})
\]
with $\urec_3^1 = \maj_3$.
We also define balanced recursive ternary majority functions
$\brec_3^h : \{0,1\}^{3^h} \mapsto \{0,1\}$ as
\begin{align*}
\brec_3^h (x_1, \cdots, x_{3^h}) = \maj_3 (
&\brec_3^{h-1}(x_1, \cdots, x_{3^{h-1}}),\\
&\brec_3^{h-1}(x_{3^{h-1}+1}, \cdots, x_{2 \cdot 3^{h-1}}),\\
&\brec_3^{h-1}(x_{2 \cdot 3^{h-1}+1}, \cdots, x_{3^h})) 
\end{align*}
with $\brec_3^1 = \maj_3$.
Through the paper, $n$ means the number of input bits.
Formula size and monotone formula size of a Boolean function $f$ are 
denoted by $L(f)$ and $L_m(f)$, respectively.
\end{definition}

Although our improvements of lower bounds seem to be slight, 
it breaks a stiff barrier (known as the certificate complexity barrier~\cite{LLS06})
of previously known proof techniques.
The best monotone upper and lower bounds of majority functions are 
$O(n^{5.3})$~\cite{Valiant84} and $\lfloor n/2 \rfloor n$~\cite{Rad97}, respectively.
In the non-monotone case, the best formula size upper and lower bounds of majority functions are
$O(n^{4.57})$~\cite{PPZ92} and $\lceil n/2  \rceil^2$ 
($=(l+1)^2$ when $n=2l+1$), respectively,
which can be proven by the classical result of
Khrapchenko~\cite{Khrapchenko71}.
In this paper, we slightly improve the non-monotone formula size lower bound
while no previously known techniques has been able to improve it since 1971.
In Section~\ref{maj_section}, we will prove 
$\frac{(l+1)^2}{1-\epsilon(l)} \leq L(\maj_{2l+1})$
where $\epsilon(l) = \frac{l^2(l+1)}{6 \cdot \binom{2l+1}{l}}$.
Here, $\binom{n}{k}$ denotes ${}_{n}C_{k}$.
Since formula size takes an integral value, it implies a $(l+1)^2 + 1$ lower bound.

It is known that the class of monotone self-dual Boolean functions is
closed under compositions (equivalently, in so-called Post's lattice~\cite{BCRV03,Post41}).
Any monotone self-dual Boolean functions can be decomposed into compositions 
of 3-bit majority functions~\cite{IK93}.
A key observation for our proofs is that a communication matrix (defined in the next section)
of a monotone self-dual Boolean function contains those of the 3-bit majority function
as its submatrices. 
Ibaraki and Kameda~\cite{IK93} developed a decomposition theory of 
monotone self-dual Boolean functions
in the context of mutual exclusions in distributed systems.
The theory has been further investigated by \cite{BI95,BIM99}.
Given a monotone self-dual Boolean function $f$,
we can decompose it as $f = \maj_3(x, f_1, (\maj_3(x, f_2, \maj_3(\cdots \maj_3(x, f_{k-1}, f_k)))))$
after decomposing $g=f(x=0)$ into a conjunction of monotone self-dual functions
$g = f_1 \wedge f_2 \wedge \cdots \wedge f_k$. 
It holds $\urec_3^h$ in its internal structure.
To determine its formula size is of particular interest because it is related 
with efficiency of the decomposition scheme.
In Section~\ref{urec_section}, we will prove
$L(\urec_3^h) = L_m(\urec_3^h) = 4h + 1$.

Balanced recursive ternary majority functions have been studied 
in several contexts~\cite{JKS03,LLS06,MO03,O04,RS08,SW86}, 
see \cite{LLS06} and \cite{RS08} for details.
Ambainis et al.~\cite{ACRSZ07} showed a quantum algorithm which evaluates 
a monotone formula of size $N$ (or called AND-OR formula) in $N^{1/2+o(1)}$ time 
even if it is not balanced.
This result implies $\brec_3^h$ can be evaluated in $O(\sqrt{5}^h)$ time by 
the quantum algorithm because we have a formula size upper bound $L_m(\brec_3^h) \leq 5^h$
as noted in~\cite{LLS06}.
Improving this result, Reichardt and Spalek~\cite{RS08} gave a quantum algorithm which
evaluates $\brec_3^h$ in $O(2^h)$ time.
From this context, seeking the true bound of the monotone formula size of 
$\brec_3^h$ is a very interesting research question.
The quantum adversary bound~\cite{LLS06} has a quite nice property written as 
$\mathbf{ADV}(f \cdot g) \geq \mathbf{ADV}(f) \cdot \mathbf{ADV}(g)$.
It directly implies a formula size lower bound $4^h \leq L(\brec_3^h)$.
In Section~\ref{brec_section}, we will prove $20 \leq L_m(\brec_3^2)$ and
$4^h + \frac{13}{36} \cdot \left( \frac{8}{3} \right)^h \leq L_m(\brec_3^h)$.
This gives a slight improvement of the lower bound
and means that the $4^h$ lower bound is at least not optimal in the monotone case.

\section{Preliminaries}
We define a total order $0 < 1$ between the two Boolean values.
For Boolean vectors $\vec{x} = (x_1, \cdots, x_n)$ and $\vec{y} = (y_1, \cdots, y_n)$,
we define $\vec{x} \leq \vec{y}$ if $x_i \leq y_i$ for all $i \in \{1, \cdots n\}$.
A Boolean function $f$ is called monotone if $\vec{x} \leq \vec{y}$ implies $f(\vec{x}) \leq f(\vec{y})$ 
for all $\vec{x}, \vec{y} \in \{ 0,1\}^n$.
For a monotone Boolean function $f$, a Boolean vector $\vec{x} \in \{ 0,1\}^n$ is called
minterm if $f(\vec{x}) =1$ and ($\vec{y} \leq \vec{x}) \wedge (\vec{x} \neq \vec{y})$ 
implies $f(\vec{y})=0$ for any $\vec{y} \in \{ 0,1\}^n$
and called maxterm if $f(\vec{x}) =0$ and ($\vec{x} \leq \vec{y}) \wedge (\vec{x} \neq \vec{y})$
implies $f(\vec{y})=1$ for any $\vec{y} \in \{ 0,1\}^n$.
Sets of all minterms and maxterms of a monotone Boolean function $f$ are 
denoted by $minT(f)$ and $maxT(f)$, respectively.
A Boolean function $f$ is called self-dual if 
$f(x_1,\cdots,x_n) = \overline{f(\overline{x}_1,\cdots,\overline{x}_n)}$ 
where $\overline{x}$ is the negation of $x$. 
Remark that, if a Boolean function $f$ is self-dual, its communication matrix (see below)
has some nice properties, e.g. $|X| = |Y|$.

A formula is a binary tree with leaves labeled by literals and 
internal nodes labeled by $\wedge$ and $\vee$. 
A literal is either a variable or the negation of a variable.
A formula is called monotone if it does not have negations.
It is known that all (monotone) Boolean functions can be represented by a (monotone) formula.
The size of a formula is its number of leaves.
We define the (monotone) formula size of a Boolean function $f$
as the size of the smallest formula computing $f$.

Karchmer and Wigderson~\cite{KW90} characterize formula size of any Boolean function 
in terms of a communication game called the Karchmer-Wigderson game.
In the game, given a Boolean function $f$, 
Alice gets an input $\vec{x}$ such that $f(\vec{x})=1$ and 
Bob gets an input $\vec{y}$ such that $f(\vec{y})=0$.
The goal of the game is to find an index $i$ such that $x_i \neq y_i$.
They also characterize monotone formula size 
by a monotone version of the Karchmer-Wigderson game.
In the monotone game, Alice gets a minterm $\vec{x}$ and Bob gets a maxterm $\vec{y}$.
The goal of the monotone game is to find an index $i$ such that $x_i = 1$ and $y_i = 0$.
The number of leaves in a best communication protocol for 
the (monotone) Karchmer-Wigderson game is equal to the (monotone) formula size of $f$.
From these characterizations, we consider communication matrices derived from the games.

\begin{definition}[Communication Matrix]
Given a Boolean function $f$, we define its communication matrix as a matrix whose rows and columns are indexed by $X=f^{-1}(1)$ and $Y=f^{-1}(0)$, respectively.
Each cell of the matrix contains indices $i$ such that $x_i \neq y_i$.
In a monotone case, given a monotone Boolean function $f$, we define its monotone communication matrix as a matrix whose rows and columns are indexed by 
$X=minT(f)$ and $Y=maxT(f)$, respectively.
Each cell of the matrix contains indices $i$ 
such that $x_i = 1$ and $y_i = 0$.
A combinatorial rectangle is a direct product $X' \times Y'$ 
where $X' \subseteq X$ and $Y' \subseteq Y$.
A combinatorial rectangle $X' \times Y'$ is called monochromatic
if every cell $(\vec{x},\vec{y}) \in X' \times Y'$ contains the same index $i$.
We call a cell singleton if it contains just one index.
\end{definition}

The minimum number of disjoint monochromatic rectangles which exactly cover all cells 
in the (monotone) communication matrix gives a lower bound for the number of leaves of 
a best communication protocol for the (monotone) Karchmer-Wigderson game. 
Thus, we obtain the following bound.

\begin{theorem}[Rectangle Bound~\cite{KW90}]
The minimum size of an exact cover by disjoint monochromatic rectangles 
for the communication matrix (or monotone communication matrix) associated 
with a Boolean function $f$ gives a lower bound of $L(f)$ (or $L_m(f)$).
\end{theorem}

\section{A Stronger Linear Programming Bound via Clique Constraints}
In this study, we devise a new technique proving formula size lower bounds based on 
the LP bound~\cite{KKN95} with clique constraints.
We assume that readers are familiar with the basics of the linear and integer programming theory.
Karchmer, Kushilevitz and Nisan~\cite{KKN95} formulate the rectangle bound 
as an integer programming problem and give its LP relaxation.
Given a (monotone) communication matrix, 
it can be written as $\min \sum_r x_r$ such that  $\sum_{r \ni c} x_r = 1$ 
for each cell $c$ in the matrix and $x_r \geq 0$ for each monochromatic rectangle $r$.
The dual problem can be written as $\max \sum_c w_c$ such that $\sum_{c \in r} w_c \leq 1$ 
for each monochromatic rectangle $r$.
Here, each variable $w_c$ is indexed by a cell $c$ in the matrix.
From the duality theorem, showing a feasible solution of the dual problem gives
a formula size lower bound.

Now, we introduce our stronger LP bound using clique constraints 
from the theory of stable set polytope.
We assume that each monochromatic rectangle is a node of a graph.
We connect two nodes by an edge if the two corresponding monochromatic rectangles intersect.
If a set of monochromatic rectangles $q$ compose a clique in the graph, 
we add a constraint $\sum_{r \in q} x_r \leq 1$ to 
the primal problem of the LP relaxation.
This constraint is valid for all integral solutions since we consider the disjoint cover problem.
That is, we can assign the value 1 to at most 1 rectangle in a clique for all integral solutions
under the condition of disjointness.
The dual problem can be written as $\max \sum_c w_c + \sum_q z_q$ such that 
$\sum_{c \in r} w_c + \sum_{q \ni r} z_q \leq 1$ for each monochromatic rectangle $r$ and
$z_q \leq 0$ for each clique $q$.
Intuitively, this formulation can be interpreted as follows.
Each cell $c$ is assigned a weight $w_c$.
The summation of weights over all cells in a monochromatic rectangle is limited to 1.
This limit is relaxed by 1 if it is contained by a clique.
Thus, the limit of the total weight for a monochromatic rectangle 
contained by $k$ distinct cliques is $k+1$.

By using clique constraints, we obtain the following matching lower bound for the formula size of
the 3-bit majority function while the original LP bound cannot prove a lower bound larger than 4.5.
In our proofs, we utilize the following property of combinatorial rectangles 
which is trivial from the definition.
If a rectangle contains two cells $(\alpha_1, \beta_1)$ and $(\alpha_2, \beta_2)$, 
it also contains both $(\alpha_1, \beta_2)$ and $(\alpha_2, \beta_1)$.
A notion of singleton cells also occupies an important role for our proofs
because there are no monochromatic rectangles which contain different kinds of singleton cells.

\begin{theorem}
\label{maj3_theorem}
$L(\maj_3) = L_m(\maj_3) = 5$
\end{theorem}
\proof
We have a monotone formula $(x_1 \wedge x_2) \vee ((x_1 \vee x_2) \wedge x_3)$ for $\maj_3$.
From the definition, $L(\maj_3) \leq L_m(\maj_3)$.
To prove $L(\maj_3) \geq 5$, we consider a communication matrix of the 3-bit majority function
whose rows and columns are restricted to minterms and maxterms, respectively.

\begin{figure}[h]
\begin{center}
\begin{tabular}{|c||c|c|c|}\hline
&100&010&001\\\hline\hline
110&2&1&1,2,3\\\hline
101&3&1,2,3&1\\\hline
011&1,2,3&3&2\\\hline
\end{tabular}
\end{center}
\caption{The Communication Matrix of $\mathbf{MAJ_3}$}
\label{3bit_fig}
\end{figure}

In the dual problem, we assign weights 1 for all singleton cells and 0 for other cells.
There are 6 singleton cells and hence the total weight is 6.
We take a clique $q$ composed of monochromatic rectangles containing two singleton cells.
It is clear that every pair of monochromatic rectangles contained by $q$ intersect at some cell.
We assign $z_q = -1$. Then, the objective function of the dual problem becomes $5 = 6-1$.

Now, we show that all constraints of the dual problem are satisfied.
First, we consider a monochromatic rectangle which contains at most one singleton cell.
In this case, the constraint is clearly satisfied because the summation of weights 
in the monochromatic rectangle is less than or equal to 1.
Then, we consider a monochromatic rectangle which contains two singleton cells.
In this case, the summation of weights in the monochromatic rectangle is 2.
However, it is contained by the clique $q$.
It implies that the limit of the total weight is relaxed by 1. Thus, the constraint is satisfied.
There are no monochromatic rectangles which contain more than 3 singleton cells 
because a rectangle which contains more than two kinds of singleton cells 
is not monochromatic.
\qed
\section{Formula Size of Majority Functions}
\label{maj_section}
In this section, we show a non-monotone formula size lower bound of majority functions
improved from the classical result of Khrapchenko~\cite{Khrapchenko71}.

\begin{theorem}
$\displaystyle L(\maj_{2l+1}) \geq \frac{(l+1)^2}{1-\epsilon(l)}$
where $\displaystyle \epsilon(l) = \frac{l^2(l+1)}{6 \cdot \binom{2l+1}{l}}$.
\end{theorem}
\proof
We consider a communication matrix of a majority function with $2l+1$ input bits
whose rows and columns are 
restricted to minterms and maxterms, respectively. 
Let $m = \binom{2l+1}{l}$, which is equal to both the number of rows and columns.
Then, the number of all cells is $m^2$.
The number of singleton cells is $(l+1) m$ and hence the number of singleton cells 
for each index is $\frac{(l+1) m}{2l+1}$.
The number of cells with 3 indices is 
$\binom{l+1}{2} \cdot l \cdot m = \frac{l^2(l+1)m}{2}$
because we can obtain a maxterm by flipping two bits of 1's to 0's and 
one bit of 0 to 1 for each minterm.

We consider $3 \times 3$ submatrices in the following way.
From $2l+1$ input bits, we fix arbitrary $2l-2$ bits and assume that they have the
same number of 0's and 1's.
Then, we consider the remaining $3$ bits.
If the $2l+1$ input bits compose a minterm, the 3 bits are 110 or 101 or 011. 
If the $2l+1$ input bits compose a maxterm, the 3 bits are 100 or 010 or 001.
Thus, we have a $3 \times 3$ submatrix, which has the same structure 
as the communication matrix of the 3-bit majority function as Figure~\ref{3bit_fig}.
The number of submatrices is $\binom{2l+1}{3} \cdot \binom{2l-2}{l-1} = \frac{l^2(l+1)m}{6}$.
Each submatrix has 6 singleton cells and 3 cells each of which has 3 indices 
corresponding to the remaining 3 bits.
Note that each cell with 3 indices in any submatrix is not contained by other submatrices.
In other words, all the $\frac{l^2(l+1)m}{2}$ cells with 3 indices are exactly partitioned into 
the $\frac{l^2(l+1)m}{6}$ submatrices.

We assign weights $a$ for all singleton cells, 0 for cells with 3 indices
and $b$ for other cells, which have more than 3 indices.
Note that there are no cells with 2 indices.
We consider $\frac{l^2(l+1)m}{6}$ clique constraints assigned weights $c$ ($\leq 0$) 
for all the $\frac{l^2(l+1)m}{6}$ submatrix. 
That is, we have a clique constraint for each submatrix 
similar to the proof of Theorem~\ref{maj3_theorem}.
More precisely, a clique associated with a submatrix is composed of
monochromatic rectangles which contain two singleton cells in the submatrix.

Then, the objective function of the dual problem is written as
\begin{equation}
\max_{a,b,c} (l+1) m \cdot a + \left( m^2 - (l+1) m - \frac{l^2(l+1) m}{2} \right) 
\cdot b + \frac{l^2(l+1) m}{6} \cdot c .
\end{equation}
Now, we fix $c=2b \leq 0$. Then, we have
\begin{equation}\label{maj_obj2}
\max_{a,b} (l+1) m \cdot a + \left( m^2 - (l+1) m - \frac{l^2(l+1) m}{6} \right) \cdot b.
\end{equation}

We assume that a monochromatic rectangle contains $k$ singleton cells and
consider all possible pairs of 2 singleton cells taken from the $k$ singleton cells.
If a pair is in the same submatrix,
the monochromatic rectangle is contained by a clique associated with the submatrix.
If a pair is not in the same submatrix,
the monochromatic rectangle contains two cells which are assigned weights $b$ because 
they have more than 3 indices.
Thus, if the following inequality is satisfied
\begin{equation}\label{maj_eq2}
k \cdot a + (k^2 - k) \cdot b \leq 1
\end{equation}
for any integer $k$ ($1 \leq k \leq \frac{(l+1)m}{2l+1}$),
all constraints of the dual problem are satisfied when $c=2b$.

We can maximize (\ref{maj_obj2}) by assuming that the inequality is saturated when
$k=\frac{m}{l+1} - \frac{l^2}{6}$ 
as it satisfies $\frac{k^2 - k}{k} = \frac{m^2 - (l+1) m - \frac{l^2(l+1) m}{6}}{(l+1) m}$.
In this case, we have 
$(\ref{maj_obj2}) = \frac{(l+1) m}{\frac{m}{l+1} - \frac{l^2}{6}}
= \frac{(l+1)^2m}{m - \frac{1}{6}l^2(l+1)} $
and obtain the lower bound.
\qed

\section{Formula Size of Unbalanced Recursive Ternary Majority Functions}
\label{urec_section}
In this section, we show the following matching bound of formula size for unbalanced
recursive ternary majority functions.
\begin{theorem}
$L(\urec_3^h) = L_m(\urec_3^h) = 4h + 1$
\end{theorem}
\proof
First, we look at the monotone formula size upper bound. 
Recall that a monotone formula of the 3-bit majority function
can be written as $(x_1 \wedge x_2) \vee ((x_1 \vee x_2) \wedge x_3)$.
The important point here is that the literal $x_3$ appears only once.
We construct $(x_{2h} \wedge x_{2h+1}) \vee ((x_{2h} \vee x_{2h+1}) \wedge x_{2h-1})$ 
and replace $x_{2h-1}$ by a monotone formula representing $\urec_3^{h-1}$.
A recursive construction yields a $4h + 1$ monotone formula for $\urec_3^h$.

Then, we show the non-monotone formula size lower bound. 
Before using clique constraints, we consider the original LP bound.
We restrict the communication matrix of $\urec_3^h$ to a submatrix $S_h$
whose rows and columns are minterms and maxterms, respectively. 
We can interpret it in the following recursive way as Figure~\ref{urec_fig}.
\begin{figure}[h]
\begin{center}
\begin{tabular}{|c||c|c|c|}\hline
     & 00 & 10 & 01 \\\hline\hline
11 & $2h, 2h+1$ & $2h+1$ & $2h$ \\\hline
01 & $2h+1$ & $T_{h-1}$ & $S_{h-1}$ \\\hline
10 & $2h$ & $S_{h-1} $ & $T_{h-1}$\\\hline
\end{tabular}
\end{center}
\caption{Recursive Structure of $S_h$ for $\urec_3^h$ ($h \geq 2$)}
\label{urec_fig}
\end{figure}

In the figure, ``11'' denotes a minterm which has 1 in the $2h$-th and 
$(2h+1)$-th bits and 0 in other $(2h-1)$ bits.
Minterms denoted by ``01'' has 0 in the $2h$-th bit and 1 in the $(2h+1)$-th bit
and other $(2h-1)$ bits of them are determined by a recursive way from
minterms of $\urec_3^{h-1}$.
Minterms denoted by ``10'' has 1 in the $2h$-th bit and 0 in the $(2h+1)$-th bit
and other $(2h-1)$ bits of them are also determined by the recursive way.
``00'', ``10'' and ``01'' denote maxterms which are similarly defined as minterms.
A submatrix $T_{h-1}$ does not contain singleton cells because all cells in
$T_{h-1}$ contains indices $\{2h,2h+1\}$ with indices of corresponding cell in $S_{h-1}$.
$S_h$ contains two $S_{h-1}$.
Thus, the number of singleton cells duplicate in each recursion.

We consider the minimum submatrix $\mathbf{ALL{-}S}_1$ in $S_h$ 
which contains all three kinds of singleton cells $\{1\}$, $\{2\}$ and $\{3\}$.
Note that $\mathbf{ALL{-}S}_1$ does not contain any other kinds of singleton cells because
it only contains cells in $S_1$ and $T_{l}$ ($2 \leq l \leq h-1$).
A submatrix $S_1$ is equivalent to a communication matrix of the 3-bit majority function.
The total number of singleton cells $\{1\}$, $\{2\}$ and $\{3\}$ is $3 \cdot 2^h$.
Both the number of rows and columns of $\mathbf{ALL{-}S}_1$ is equal to $3 \cdot 2^{h-1}$
because $S_1$'s duplicate $(h-1)$-times and does not have any common rows and columns.
Hence, the number of all cells in $\mathbf{ALL{-}S}_1$ is $9 \cdot 4^{h-1}$.
We assign weights $a$ for all singleton cells in $\mathbf{ALL{-}S}_1$ 
and weights $b$ for all other cells in $\mathbf{ALL{-}S}_1$.
Then, the total weight of all cells in $\mathbf{ALL{-}S}_1$ is written as follows:
\begin{equation}
\label{urec_obj}
\max_{a,b} 3 \cdot 2^h \cdot a + \left( 9 \cdot 4^{h-1} - 3 \cdot 2^h \right) \cdot b .
\end{equation}
We consider constraints of the dual problem as $k \cdot a + (k^2 - k) \cdot b \leq 1$
for all integer $k$ ($1 \leq k \leq 2^h$).
We assume this inequality is saturated if and only if $k= 3 \cdot 2^{h-2}$.
Then, we get $a = \frac{24 \cdot 2^h - 16}{9 \cdot 4^h}$ and $b = - \frac{16}{9 \cdot 4^h}$.
In this case, $(\ref{urec_obj}) = 4$.

Next, we consider singleton cells $\{2l\}$ and $\{2l+1\}$ ($2 \leq l \leq h$).
We partition singleton cells $\{2l\}$ into two sets named 
vertical cells $X_{2l}$ and horizontal cells $Y_{2l}$ which are in
(10,00) and (11,01) of each $S_l$ in $S_h$, respectively.
Similarly, we partition singleton cells $\{2l+1\}$ into two sets named 
vertical cells $X_{2l+1}$ and horizontal cells $Y_{2l+1}$ which are in
(01,00) and (11,10) of each $S_l$ in $S_h$, respectively.
We restrict these sets to the minimum subsets 
$X'_{2l} \subset X_{2l}$, $X'_{2l+1} \subset X_{2l+1}$, 
$Y'_{2l} \subset Y_{2l}$ and $Y'_{2l+1} \subset Y_{2l+1}$ 
so as to satisfy the following condition:
If a monochromatic rectangle contains all cells in $X'_{2l} \cup X'_{2l+1} \cup Y'_{2l} \cup Y'_{2l+1}$, 
it also contains all cells in $\mathbf{ALL{-}S}_1$.
Note that rows and columns of singleton cells $\{2l\}$ and $\{2l+1\}$ dominate 
those of singleton cells $\{1\}$, $\{2\}$ and $\{3\}$.
So, we have $|X'_{2l}| = |X'_{2l+1}| = |Y'_{2l}| = |Y'_{2l+1}| = 3 \cdot 2^{h-2}$.
We assign weights $\frac{1}{3\cdot 2^{h-2}}$ for all singleton cells 
in $X'_{2l} \cup X'_{2l+1} \cup Y'_{2l} \cup Y'_{2l+1}$ and
0 for other cells at (11,00) of each $S_l$ and cells outside $\mathbf{ALL{-}S}_1$.
A monochromatic rectangle which contains $x$ cells in $X'_{2l}$ and $y$ in from $Y'_{2l}$ also contains
$x \cdot y$ cells in $\mathbf{ALL{-}S}_1$ which are assigned weights $b$.
The same thing is true for the case of $X'_{2l+1}$ and $Y'_{2l+1}$.
Because we have
\begin{equation}
\label{urec_eq}
(x+y) \cdot \frac{4}{3\cdot 2^{h}} - x y \cdot \frac{16}{9 \cdot 4^h} \leq 1 
\end{equation}
for all $0 \leq x,y \leq 3\cdot 2^{h-2}$, all constraints of the dual problem are satisfied.
The total weight of singleton cells $\{2l\}$ and $\{2l+1\}$ is 4.
So, the total weight of all cells in $S_h$ now becomes $4h$.

Now, we incorporate clique constraints.
The number of $S_1$ in $S_h$ is $2^{h-1}$.
We change weights of all non-singleton cells in submatrices $S_1$ from $b$ to $0$.
On behalf of them, we add a clique constraint for each $S_1$ in $S_h$.
Then, $(\ref{urec_obj})$ becomes
\begin{equation}
\label{urec_obj2}
\max_{a,b,c} 3 \cdot 2^h \cdot a 
+ \left( 9 \cdot 4^{h-1} - 3 \cdot 2^h - 3 \cdot 2^{h-1} \right) \cdot b + 2^{h-1} \cdot c.
\end{equation}
where $c$ is a weight assigned for each clique constraint.
If we take $a = \frac{24 \cdot 2^h - 16}{9 \cdot 4^h}$, $b = - \frac{16}{9 \cdot 4^h}$
and $c=2b$, all constraints of the dual problem are satisfied and 
$(\ref{urec_obj2})=4 + \frac{8}{9} \cdot 2^{-h}$.
Consequently, the total weight is $4h + \frac{8}{9} \cdot 2^{-h}$. 
Since formula size must be an integer, we have shown the theorem.
\qed
\section{Monotone Formula Size of Balanced Recursive Ternary Majority Functions}
\label{brec_section}
In this section, we show monotone formula size lower bounds 
of balanced recursive ternary majority functions.
For this purpose, we consider rank constraints, which are generalizations of clique constraints.
Similarly to the case of clique constraints, 
we consider a graph composed of monochromatic rectangles and its induced subgraph $g$.
We consider a constraint $\sum_{r \in g} x_r \leq \alpha(g)$ 
where $\alpha(g)$ is the stability number of $g$.
This constraint is valid because we can assign 1 at most $\alpha(g)$ rectangles in $g$
for any integral solution.
The dual problem can be written as $\max \sum_c w_c + \sum_q z_q + \sum_g \alpha(g) z_g$ 
such that $\sum_{c \in r} w_c + \sum_{q \ni r} z_q + \sum_{g \ni r} z_g \leq 1$ 
for each monochromatic rectangle $r$,
$z_q \leq 0$ for each clique $q$ and $z_g \leq 0$ for each subgraph $g$.

First, we consider the case of height 2.
By using clique constraints and rank constraints, 
we prove the following improved monotone formula size lower bound
while we know that the original LP bound cannot prove a lower bound larger than 16.5.
\begin{theorem}
\label{brec2_theorem}
$L_m(\brec_3^2) \geq 20$
\end{theorem}
\proof
There are 27 minterms and 27 maxterms for the recursive ternary majority function of height 2.
Among them, we choose the following 9 minterms
\begin{center}
\begin{tabular}{ccc}
110,110,000 & 101,101,000 & 011,011,000\\
110,000,110 & 101,000,101 & 011,000,011\\
000,110,110 & 000,101,101 & 000,011,011
\end{tabular}
\end{center}
and 9 maxterms
\begin{center}
\begin{tabular}{ccc}
111,100,100 & 111,010,010 & 111,001,001\\
100,111,100 & 010,111,010 & 001,111,001\\
100,100,111 & 010,010,111 & 001,001,111.
\end{tabular}
\end{center}
From these 9 minterms and 9 maxterms, a submatrix of the communication matrix 
can be described as Figure~\ref{brec_fig1}.
In the figure, we abbreviate a minterm e.g. 101,101,000 by 110 and 101,
which represent the second level and the first level structure of the 9 bits, respectively.
Notice that all minterms which we choose have the same structure 
in all 3-bits minterm blocks at the first level.
The same thing is true for all 9 maxterms.

\begin{figure}[h]
\begin{center}
\begin{tabular}{|c|c||c|c|c||c|c|c||c|c|c|}\hline
&& \multicolumn{3}{|c||}{100} & \multicolumn{3}{|c||}{010} & \multicolumn{3}{|c|}{001}\\\hline
&& 100 & 010 & 001 & 100 & 010 & 001 & 100 & 010 & 001\\\hline\hline
     &110&5&4&4,5&2&1&1,2&2,5&1,4&1,2,4,5\\\cline{2-11}
110&101&6&4,6&4&3&1,3&1&3,6&1,3,4,6&1,4\\\cline{2-11}
     &011&5,6&6&5&2,3&3&2&2,3,5,6&3,6&2,5\\\hline\hline
     &110&8&7&7,8&2,8&1,7&1,2,7,8&2&1&1,2\\\cline{2-11}
101&101&9&7,9&7&3,9&1,3,7,9&1,7&3&1,3&1\\\cline{2-11}
     &011&8,9&9&8&2,3,8,9&3,9&2,8&2,3&3&2\\\hline\hline
     &110&5,8&4,7&4,5,7,8&8&7&7,8&5&4&4,5\\\cline{2-11}
011&101&6,9&4,6,7,9&4,7&9&7,9&7&6&4,6&4\\\cline{2-11}
     &011&5,6,8,9&6,9&5,8&8,9&9&8&5,6&6&5\\\hline
\end{tabular}
\end{center}
\caption{A Submatrix of the Communication Matrix for $\brec_3^2$}
\label{brec_fig1}
\end{figure}
\begin{figure}[h]
\begin{center}
\begin{tabular}{|c|c||c|c|c||c|c|c||c|c|c|}\hline
&& \multicolumn{3}{|c||}{100} & \multicolumn{3}{|c||}{010} & \multicolumn{3}{|c|}{001}\\\hline
&& 100 & 010 & 001 & 100 & 010 & 001 & 100 & 010 & 001\\\hline\hline
     &110&1&2&3&4&5&6&7&8&9\\\cline{2-11}
110&101&10&11&12&13&14&15&16&17&18\\\cline{2-11}
     &011&19&20&21&22&23&24&25&26&27\\\hline\hline
     &110&28&29&30&31&32&33&34&35&36\\\cline{2-11}
101&101&37&38&39&40&41&42&43&44&45\\\cline{2-11}
     &011&46&47&48&49&50&51&52&53&54\\\hline\hline
     &110&55&56&57&58&59&60&61&62&63\\\cline{2-11}
011&101&64&65&66&67&68&69&70&71&72\\\cline{2-11}
     &011&73&74&75&76&77&78&79&80&81\\\hline
\end{tabular}
\end{center}
\caption{Serial Numbers for 81 cells of the Submatrix}
\label{brec_fig2}
\end{figure}

To describe 12 cliques $q_1, \cdots, q_{12}$ and a induced subgraph $g$ whose stability number is 4, we give serial numbers for 81 cells as Figure~\ref{brec_fig2}.
We take the following 12 cliques each of which consists of 3 pairs of 2 singleton cells:
\begin{center}
\{ (5, 15), (4, 24), (13, 23) \},
\{ (35, 45), (34, 54), (43, 53) \},\\
\{ (2, 12), (1, 21), (10, 20) \},
\{ (62, 72), (61, 81), (70, 80) \},\\
\{ (29, 39), (28, 48), (37, 47) \},
\{ (59, 69), (58, 78), (67, 77) \},\\
\{ (5, 35), (2, 62), (29, 59) \},
\{ (15, 45), (12, 72), (39, 69) \},\\
\{ (4, 34), (1, 61), (28, 58) \},
\{ (24, 54), (21, 81) (48, 78) \},\\
\{ (13, 43), (10, 70), (37, 67) \},
\{ (23, 53), (20, 80), (47, 77) \}.
\end{center}
For each combination of 3 pairs, it is easy to verify that rectangles 
each of which contains both of two singleton cells from one of the 3 pairs 
compose a clique.

Next, we consider the following 18 pairs of singleton cells which induce the subgraph $g$:
\begin{center}
(5, 45), (15, 35), (4, 54), (24, 34), (13, 53), (23, 43),
(2, 72), (12, 62), (1, 81),\\ (21, 61), (10, 80), (20, 70),
(29, 69), (39, 59), (28, 78), (48, 58), (37, 77), (47, 67).
\end{center}
If a rectangle contain both of two singleton cells from one of 18 pairs, it also contains 2 cells from 9 cells 
\{ 9, 17, 25, 33, 41, 49, 57, 65, 73 \}.
Thus, we can choose at most 4 pairs without conflicts from 18 pairs.
It implies that the stability number of $g$ is 4.

Notice that all these 12 cliques and the subgraph cover all pairs of two singleton cells which have the same index.
We assign 1 for all 36 singleton cells in this submatrix and 0 for other cells.
We take $z_{q_1} = \cdots = z_{q_{12}} = z_{g} = -1$.
Then, the objective value of the dual problem becomes $36 - 12 - 4 = 20$.
If a rectangle contains at most one singleton cell, the constraint of the dual problem is trivially satisfied.
If a rectangle contains $k$ $(2 \leq k \leq 4)$ singleton cells, it is covered by $k-1$ cliques or $k-2$ cliques plus the subgraph $g$.
So, the constraint is also satisfied.
As a consequence, we obtain the formula size lower bound.
\qed

Note that we need a much more complicated argument to look at the non-monotone case, 
which we do not investigate in this paper, 
because singleton cells in the monotone communication matrix are not singleton in 
the non-monotone communication matrix. 

In the general monotone case, we can prove a slightly better lower bound than 
the quantum adversary bound~\cite{LLS06}, which shows a $4^h$ lower bound.
\begin{theorem}
\label{brec_theorem}
$L_m(\brec_3^h) \geq 4^h + \frac{13}{36} \cdot 
\left( \frac{8}{3} \right)^h $ $(h \geq 2)$
\end{theorem}
\proof
First, we choose $3^h$ minterms and $3^h$ maxterms 
from $3^h$ input bits of $\brec_3^h$ so as to have the same structure 
in the 1st, 2nd, $\cdots$ and $h$-th levels in the following sense.
In the $l$-th level, we have $3^{h-l}$ bits which are recursively constructed from lower levels 
in the following way.
We partition $3^l$ bits into $3^{l-1}$ blocks each of which contains consecutive 3 bits.
For each block of 3 bits, we replace them into 1 bit 
which is the output of $\maj_3$ with the 3 bits.
Then, we get $3^{h-(l+1)}$ bits.
We have $3^h$ bits as input bits in the first level
and can construct them for each level by induction.
If all of $3^{l-1}$ blocks have the same 3 bits except 000 and 111 in the case of minterms and maxterms, respectively, we call that they have the same structure in the $l$-the level.
There are $3^h$ minterms and $3^h$ maxterms because we have 3 choices in each level.
We consider the submatrix whose rows and columns are composed of these $3^h$ minterms and 
$3^h$ maxterms, respectively.

From another viewpoint, we can interpret it as a recursively construction of 
the submatrix $S_h$ of the communication matrix of $\brec_3^h$ as follows.
We define $S_h(k)$ $(k=1,2,3)$ as a matrix such that
some cell of $S_h(k)$ contains an index $(k-1) \cdot 3^h + i$ if and only if
the corresponding cell of $S_h$ contains an index $i$.
By induction, we can see that the number of all cells and singleton cells in 
$S_h$ is $9^h$ and $6^h$, respectively.
Singleton cells of each index from $3^h$ bits in $S_h$ is $2^h$.
Indices of cells in $T_h(1,2)$, $T_h(2,3)$ and $T_h(2,3)$ in Figure~\ref{brec_fig3} can be determined 
from the property of combinatorial rectangles, but we do not go to the details because 
we will assign the same weight for all these cells in each level.

\begin{figure}[h]
\begin{center}
\begin{tabular}{|c||c|c|c|}\hline
&100&010&001\\\hline\hline
110& $S_{h-1}(2)$ & $S_{h-1}(1)$ & $T_{h-1}(1,2)$ \\\hline
101& $S_{h-1}(3)$ & $T_{h-1}(2,3)$ & $S_{h-1}(1)$ \\\hline
011& $T_{h-1}(2,3)$ & $S_{h-1}(3)$ & $S_{h-1}(2)$\\\hline
\end{tabular}
\end{center}
\caption{Recursive Structure of $S_h$  for $\brec_3^h$ ($h \geq 2$)}
\label{brec_fig3}
\end{figure}

Before using clique and rank constraints, we consider the original LP bound.
We assign weights $a$ for all singleton cells, $b$ for other cells in the submatrix and 0 
for all cells in the outside of the submatrix.
Then, the objective value of the dual problem is written as
\begin{equation}
\label{obj}
\max_{a,b} 6^h \cdot a + (9^h - 6^h) \cdot b .
\end{equation}
If a rectangle contains $k$ singleton cells, it also contains at least $k^2 - k$ cells which are not singleton.
Thus, if $k \cdot a + (k^2 - k) \cdot b \leq 1$ is satisfied
for all integer $k$ ($1 \leq k \leq 2^h$), then all constraints of the dual problem are also satisfied.
We assume that the inequality is saturated if and only if $k = (3/2)^h$.
Then, we get $a = \frac{2 \cdot 6^h - 4^h}{9^h}$ and 
$b = - \frac{4^h}{9^h}$.
In this case, we have $(\ref{obj}) = 4^h$.

Now, we incorporate clique and rank constraints.
We change weights of all cells except singleton cells in all $S_2$'s 
in the second level from $b$ to $0$.
Then, we add 12 clique constraints and a rank constraint for each $S_2$ in the second level 
by following the way of Theorem~\ref{brec2_theorem}.
Let $c$ and $d$ be values assigned for every clique and rank constraints, respectively.
Then, the objective value of the dual problem is
\begin{equation}
\label{obj2}
\max_{a,b,c,d} 6^h \cdot a + (9^h - 81 \cdot 6^{h-2}) \cdot b + 12 \cdot 6^{h-2} \cdot c + 4 \cdot 6^{h-2} \cdot d .
\end{equation}
If we take $c=d=2b$, then we have
$(\ref{obj2}) = 6^h \cdot a + (9^h - 49 \cdot 6^{h-2}) \cdot b 
= 4^h + \frac{13}{36} \cdot \left( \frac{8}{3} \right)^h $.
Since all weights which are changed from $b$ to $0$ are exactly compensated by clique and rank constraints,
all constraints of the dual problem are satisfied.
\qed

We do not exhaust the potential of our new method and 
have possibilities to improve the lower bound.
For example, we can improve the lower bound as
$4^h + c \cdot \left( \frac{8}{3} \right)^h $ for some constant $c$
by further detailed analysis in constantly higher levels.

\section{Conclusions}
In this paper, we devised the new technique proving formula size lower bounds and showed
improved formula size lower bounds of some families of monotone self-dual Boolean functions 
such as majority functions, unbalanced and balanced recursive ternary majority functions.
We hope that our method will be able to improve formula size lower bounds for
any monotone self-dual Boolean function and even much broader classes of Boolean functions.
Whether our technique (or its extensions) can break the $4 n^2$ barrier and
improve the best formula size lower bound remains open.

\section*{Acknowledgment}
\noindent The author is grateful to Norie Fu for cooperating computational experiments,
Troy Lee for suggesting to look at compositions of Boolean functions,
Patric \"{O}sterg{\aa}rd for useful information about the exact cover problem, and
Kazuhisa Makino for helpful discussion.
This research is supported by Research Fellowship for Young Scientists 
from Japan Society for the Promotion of Science (JSPS) and Grant-in-Aid for JSPS Fellows.

\bibliographystyle{abbrv}
\bibliography{stacs2009}

\begin{thebibliography}{10}

\bibitem{Ambainis02}
A.~Ambainis.
\newblock Quantum lower bounds by quantum arguments.
\newblock {\em Journal of Computer and System Sciences}, 64(4):750--767, 2002.

\bibitem{ACRSZ07}
A.~Ambainis, A.~M. Childs, B.~Reichardt, R.~\v{S}palek, and S.~Zhang.
\newblock Any {AND}-{OR} formula of size {N} can be evaluated in time
  ${N}^{1/2+o(1)}$ on a quantum computer.
\newblock In {\em Proceedings of the 48th Annual IEEE Symposium on Foundations
  of Computer Science}, pages 363--372, 2007.

\bibitem{BI95}
J.~C. Bioch and T.~Ibaraki.
\newblock Decompositions of positive self-dual boolean functions.
\newblock {\em Discrete Mathematics}, 140(1-3):23--46, 1995.

\bibitem{BIM99}
J.~C. Bioch, T.~Ibaraki, and K.~Makino.
\newblock Minimum self-dual decompositions of positive dual-minor boolean
  functions.
\newblock {\em Discrete Applied Mathematics}, 96-97:307--326, 1999.

\bibitem{BCRV03}
E.~B{\"o}hler, N.~Creignou, S.~Reith, and H.~Vollmer.
\newblock Playing with boolean blocks, part {I}: Post's lattice with
  applications to complexity theory.
\newblock {\em ACM SIGACT News}, 34(4):38--52, 2003.

\bibitem{GLS88}
M.~Gr{\"o}tschel, L.~Lov{\'a}sz, and A.~Schrijver.
\newblock {\em Geometric Algorithms and Combinatorial Optimization}.
\newblock Springer, 1988.

\bibitem{Hastad98}
J.~H{\aa}stad.
\newblock The shrinkage exponent of {De~Morgan} formulas is~2.
\newblock {\em SIAM Journal on Computing}, 27(1):48--64, Feb. 1998.

\bibitem{HLS07}
P.~H{\o}yer, T.~Lee, and R.~\v{S}palek.
\newblock Negative weights make adversaries stronger.
\newblock In {\em Proceedings of the 39th Annual {ACM} Symposium on Theory of
  Computing}, pages 526--535, 2007.

\bibitem{IK93}
T.~Ibaraki and T.~Kameda.
\newblock A theory of coteries: Mutual exclusion in distributed systems.
\newblock {\em IEEE Transactions on Parallel and Distributed Computing},
  PDS-4(7):779--794, July 1993.

\bibitem{JKS03}
T.~S. Jayram, R.~Kumar, and D.~Sivakumar.
\newblock Two applications of information complexity.
\newblock In {\em Proceedings of the 35th Annual ACM Symposium on Theory of
  Computing}, pages 673--682, 2003.

\bibitem{KKN95}
M.~Karchmer, E.~Kushilevitz, and N.~Nisan.
\newblock Fractional covers and communication complexity.
\newblock {\em SIAM Journal on Discrete Mathematics}, 8(1):76--92, Feb. 1995.

\bibitem{KW90}
M.~Karchmer and A.~Wigderson.
\newblock Monotone circuits for connectivity require super-logarithmic depth.
\newblock {\em SIAM Journal on Discrete Mathematics}, 3(2):255--265, May 1990.

\bibitem{Khrapchenko71}
V.~M. Khrapchenko.
\newblock Complexity of the realization of a linear function in the case of
  $\pi$-circuits.
\newblock {\em Mathematical Notes}, 9:21--23, 1971.

\bibitem{Koutsoupias93}
E.~Koutsoupias.
\newblock Improvements on {Khrapchenko}'s theorem.
\newblock {\em Theoretical Computer Science}, 116(2):399--403, Aug. 1993.

\bibitem{LLS06}
S.~Laplante, T.~Lee, and M.~Szegedy.
\newblock The quantum adversary method and classical formula size lower bounds.
\newblock {\em Computational Complexity}, 15(2):163--196, 2006.

\bibitem{Lee07}
T.~Lee.
\newblock A new rank technique for formula size lower bounds.
\newblock In {\em Proceedings of the 24th Annual Symposium on Theoretical
  Aspects of Computer Science}, volume 4393 of {\em Lecture Notes in Computer
  Science}, pages 145--156. Springer, 2007.

\bibitem{MO03}
E.~Mossel and R.~O'Donnell.
\newblock On the noise sensitivity of monotone functions.
\newblock {\em Random Structures and Algorithms}, 23(3):333--350, 2003.

\bibitem{O04}
R.~O'Donnell.
\newblock Hardness amplification within ${NP}$.
\newblock {\em Journal of Computer and System Sciences}, 69(1):68--94, 2004.

\bibitem{Padberg73}
M.~Padberg.
\newblock On the facial structure of the set packing polyhedra.
\newblock {\em Mathematical Programming}, 5:199--215, 1973.

\bibitem{PPZ92}
M.~S. Paterson, N.~Pippenger, and U.~Zwick.
\newblock Optimal carry save networks.
\newblock In {\em Boolean function complexity}, volume 169 of {\em London
  Mathematical Society Lecture Note Series}, pages 174--201. Cambridge
  University Press, 1992.

\bibitem{Post41}
E.~L. Post.
\newblock {\em The two-valued iterative systems of mathematical logic},
  volume~5 of {\em Annals Mathematical Studies}.
\newblock Princeton University Press, 1941.

\bibitem{Rad97}
J.~Radhakrishnan.
\newblock Better lower bounds for monotone threshold formulas.
\newblock {\em Journal of Computer and System Sciences}, 54(2):221--226, Apr.
  1997.

\bibitem{RS08}
B.~Reichardt and R.~Spalek.
\newblock Span-program-based quantum algorithm for evaluating formulas.
\newblock In {\em Proceedings of the 40th Annual {ACM} Symposium on Theory of
  Computing}, pages 103--112, 2008.

\bibitem{SW86}
M.~E. Saks and A.~Wigderson.
\newblock Probabilistic boolean decision trees and the complexity of evaluating
  game trees.
\newblock In {\em Proceedings of the 27th Annual IEEE Symposium on Foundations
  of Computer Science}, pages 29--38, 1986.

\bibitem{Valiant84}
L.~G. Valiant.
\newblock Short monotone formulae for the majority function.
\newblock {\em Journal of Algorithms}, 5(3):363--366, Sept. 1984.

\end{thebibliography}
\vspace*{-1.5cm}
\end{document}